\documentclass[10pt,letterpaper]{article}
\usepackage[top=0.85in,left=1.2in,footskip=0.75in]{geometry}

\usepackage{changepage}

\usepackage[utf8x]{inputenc}

\usepackage{textcomp,marvosym}

\usepackage{fixltx2e}

\usepackage{amsmath,amssymb}

\usepackage{cite}

\usepackage{nameref,hyperref}

\usepackage{microtype}
\DisableLigatures[f]{encoding = *, family = * }

\usepackage[aboveskip=1pt,labelfont=bf,labelsep=period,justification=raggedright,singlelinecheck=off]{caption}

\makeatletter
\renewcommand{\@biblabel}[1]{\quad#1.}
\makeatother

\date{}

\usepackage{lastpage,fancyhdr,graphicx}
\usepackage{epstopdf}
\fancyheadoffset[L]{2.25in}
\fancyfootoffset[L]{2.25in}
\usepackage[gen]{eurosym}
\usepackage{subfig}

\begin{document}
\vspace*{0.2in}

\begin{flushleft}
{\Large
\textbf\newline{A 3D Printed Toolbox for Opto-Mechanical
Components} % Please use "title case" (capitalize all terms in the title except conjunctions, prepositions, and articles).
}
\newline
\\
Luis José Salazar-Serrano\textsuperscript{1,2,*},
Juan P. Torres\textsuperscript{2,3},
Alejandra Valencia\textsuperscript{1}
\\
\bigskip
\textbf{1} Quantum Optics Laboratory, Universidad de los Andes, AA 4976, Bogotá, Colombia
\\
\textbf{2} ICFO-Institut de Ciencies Fotoniques, Mediterranean Technology Park, 08860 Castelldefels (Barcelona), Spain
\\
\textbf{3} Dep. of Signal Theory and Communications, Universitat Politecnica de Catalunya, 08034 Barcelona, Spain
\\
\bigskip

% Insert additional author notes using the symbols described below. Insert symbol callouts after author names as necessary.
% 
% Remove or comment out the author notes below if they aren't used.
%
% Primary Equal Contribution Note
%%\Yinyang These authors contributed equally to this work.

% Additional Equal Contribution Note
% Also use this double-dagger symbol for special authorship notes, such as senior authorship.
%%\ddag These authors also contributed equally to this work.

%%\textcurrency Current Address: Dept/Program/Center, Institution Name, City, State, Country % change symbol to "\textcurrency a" if more than one current address note
% \textcurrency b Insert second current address 
% \textcurrency c Insert third current address

%%\dag Deceased

% Group/Consortium Author Note
%%\textpilcrow Membership list can be found in the Acknowledgments section.

% Use the asterisk to denote corresponding authorship and provide email address in note below.
* luis-jose.salazar@icfo.es

\end{flushleft}
\section*{Abstract}
Nowadays is very common to find headlines in the media where it is stated that 3D printing is a technology called to change our lives in the near future~\cite{Economist}. For many authors, we are living in times of a third industrial revolution~\cite{Berman,Pearce2013a}. Howerver, we are currently in a stage of development where the use of 3D printing is advantageous over other manufacturing technologies only in rare scenarios. Fortunately, scientific research is one of them~\cite{Pearce_book}. Here we present the development of a set of opto-mechanical components that can be built easily using a 3D printer based on Fused Filament Fabrication (FFF) and parts that can be found on any hardware store. The components of the set presented here are highly customizable, low-cost, require a short time to be fabricated and offer a performance that compares favorably with respect to low-end commercial alternatives. 

% Please keep the Author Summary between 150 and 200 words
% Use first person. PLOS ONE authors please skip this step. 
% Author Summary not valid for PLOS ONE submissions.   
%\section*{Author Summary}

%*** \linenumbers ***

% Use "Eq" instead of "Equation" for equation citations.
\section*{Introduction}
3D printing technology is evolving continuously in various directions. The development of scientific tools is one of the areas that is more rapidly growing since it is opening the possibility of using state-of-the-art scientific equipment at a fraction of the cost with respect to commercial alternatives~\cite{PearceScience, Baden, Pearce2013b, Pearce2014}.

The implementation of opto-mechanical components using 3D printing has a direct impact on the photonics community~\cite{Zavattieri,Piller}. Researchers are no longer constrained to work with available commercial elements and therefore their experimental setups can be more versatile. Thanks to the fact that the fabrication time is minimal, the components can evolve very fast from the researchers experience and that evolution process can be a creative way to engage young researchers in photonics. Moreover, the components 3D printed can be considered as prototypes and thus large manufacturers can take this ideas and build better equipment. 

Additionally, since the 3D printers based on Fused Filament Fabrication (FFF) are becoming more affordable the opto-mechanical components of our set can be fabricated practically in any location. It is important to highlight that since the components fabricated have a similar performance than low-end commercial alternatives in terms of stability and robustness, the approach presented here makes photonics more accessible to industry and academia both in developed and developing countries, enlarging significantly the size of the community interested in performing experiments in photonics and related fields~\cite{Euler}. In economically developed countries, the ideas presented here constitutes a way to reduce increasingly rising costs, and in developing countries, allows to overcome funding restrictions and large lead times due to customs and administrative procedures~\cite{Zavattieri}. 

The toolbox presented in this letter is composed of components highly customizable, low-cost, that require a short time to be fabricated, offer a performance that compares favourably with respect to low-end commercial alternatives, that are also aimed at complementing other component libraries already available on internet~\cite{Pearce2013b, Baden}.

\section*{Materials and Methods}
\begin{figure}[!h]
\centering
\subfloat[]
{\label{fig:figure1a}\includegraphics[width=.3\textwidth]{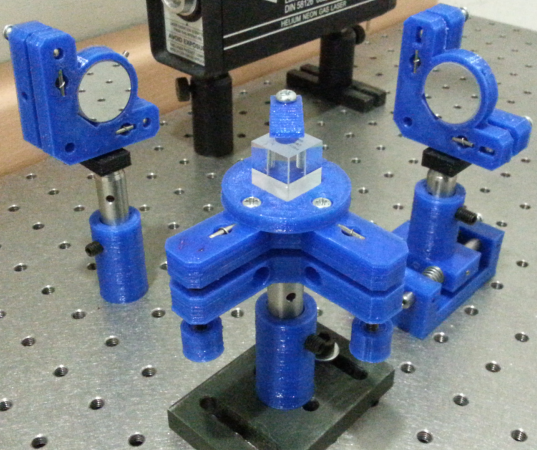}} \quad
\subfloat[]
{\label{fig:figure1b}\includegraphics[width=.3\textwidth]{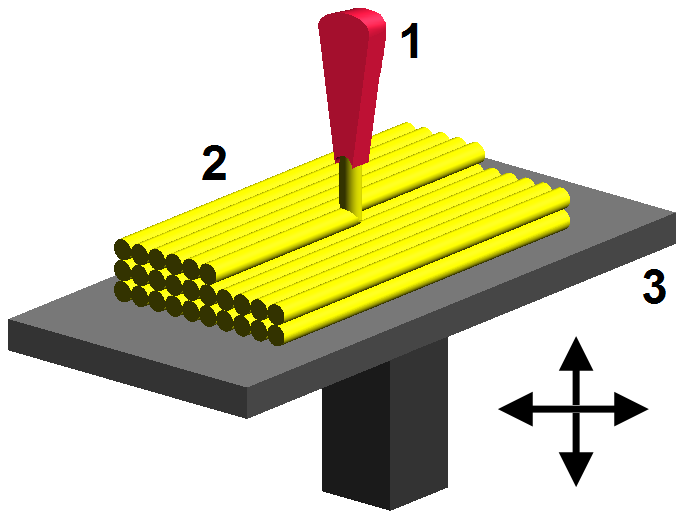}} \quad
\subfloat[]
{\label{fig:figure1c}\includegraphics[width=.3\textwidth]{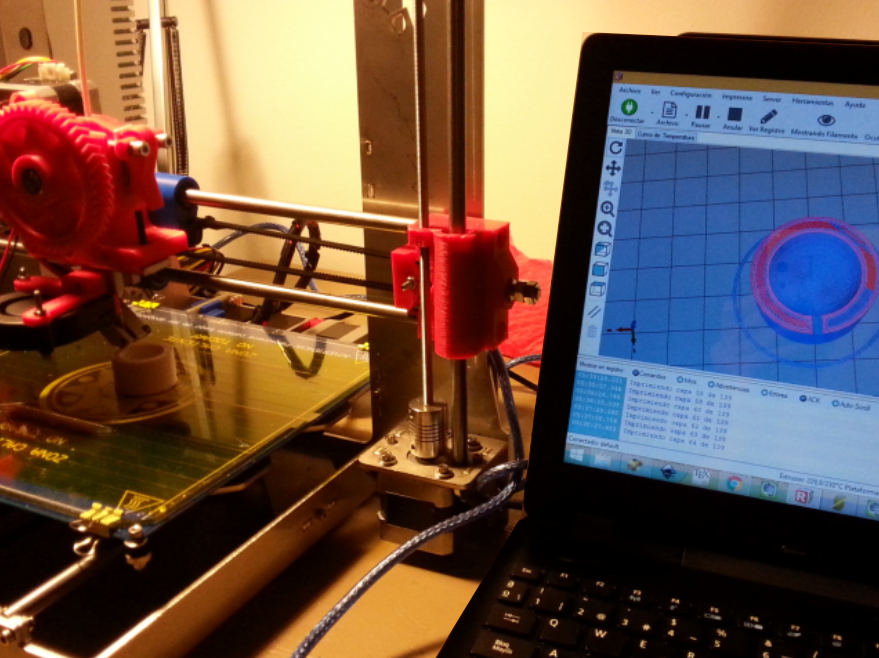}} \\
\caption{{\bf Set of opto-mechanical components printed.}
(A) A Michelson-Morley interferometer implemented with opto-mechanical components made in plastic. (B) Fused Filament Fabrication (FFF), a method of rapid prototyping: 1. Nozzle ejecting molten material (plastic), 2. Deposited material (modelled part), 3. Controlled movable table. (C) Printing a component using a Prusa-Tairona printer.}
\label{fig:figure1}
\end{figure}

We have designed and fabricated a set of opto-mechanical components that are essential in any optics laboratory either for research or teaching. For the sake of illustration we have fabricated all the opto-mechanical components to construct a Michelson-Morley interferometer, since it is an experimental setup that requires a very diverse set of opto-mechanical components for its implementation. In figure \ref{fig:figure1}A we present some of the components fabricated, such as \textbf{kinematic mounts}, used to set the tip and tilt of mirrors or lenses, \textbf{translation stages}, used to set the position of a component along a single axis with high precision, \textbf{kinematic platforms}, used to support and set the position of components such as prisms or beam splitters. In addition we have also built an \textbf{integrating sphere}, used to measure the power of a light source and other basic components such as \textbf{post holders} and \textbf{post clamps}. All the plastic parts were printed using a Prusa-Tairona~\cite{PrusaTayrona} 3D printer that cost $\approx 400\euro{}$ and is manufactured in Colombia.

The complete scheme of fabrication used to build the opto-mechanical components consists of two parts. In the first, the main elements are built in plastic using a standard 3D fabrication process that can be described as follows: to start, a 3D model of the component is designed using a CAD software for instance OpenSCAD, Blender, Solid Works, or can be downloaded from a digital design repository like Thingiverse if the model has been already designed. For our purposes, all the plastic components were designed using OpenSCAD, an open-source, script-based software that generates 3D models by combining (adding or subtracting) primitive shapes such as cylinders, spheres and cubes, that is very easy to use. Once the model is available, it is further converted using another software, like slicer or cura, into printing instructions for the 3D printer. In our case, the Prusa Tayrona printer uses the programs Repetier Host and Slicer to generate the instructions (gcode) and print the piece, respectively. Afterwards, the component is fabricated using an additive manufacturing technology known as Fused Filament Fabrication (FFF) in which a filament of PLA (Polylactic Acid, a non-toxic, biodegradable thermoplastic polymer made from plant-based resources such as corn starch or sugar cane) is heated and then extruded through a hot nozzle. The hot plastic is deposited layer by layer following a given pattern so that each layer binds with the layer below to build a solid object. A moving platform or moving nozzle determines the position of the hot filament and thus the shape of the solid object printed (figures \ref{fig:figure1}B and \ref{fig:figure1}C). 

\begin{figure}[!h]
\centering
\subfloat[]
{\label{fig:figure1a}\includegraphics[width=.3\textwidth]{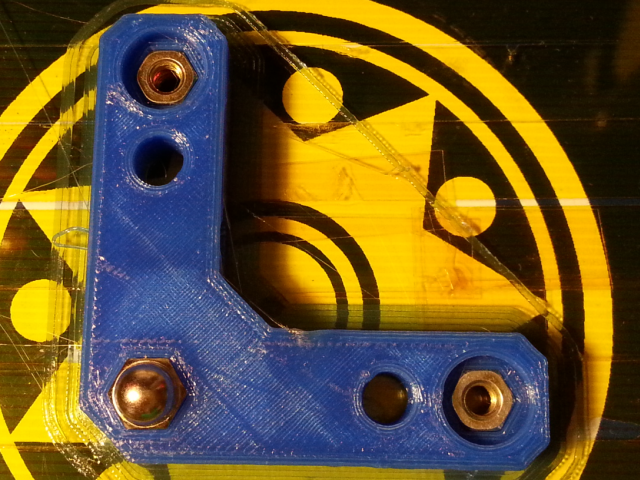}} \quad
\subfloat[]
{\label{fig:figure1b}\includegraphics[width=.3\textwidth]{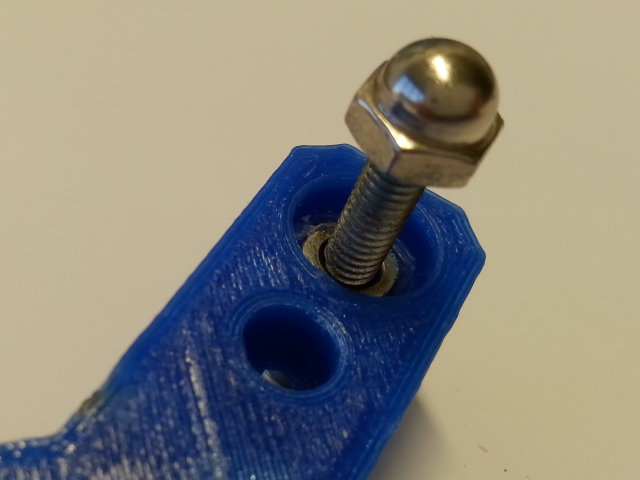}} \quad
\subfloat[]
{\label{fig:figure1c}\includegraphics[width=.3\textwidth]{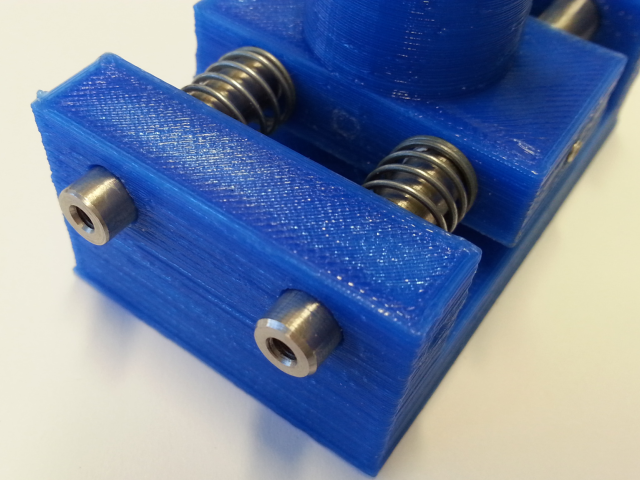}} \\
\caption{{\bf Mechanism Implementation.}
(A) Nuts of different shapes embedded on the plastic immediately after fabrication. (B) Drive screw mechanism implemented using a nut embedded in the plastic and a screw. The precision in the displacement is limited by the nut's thread. (C) Linear bearings can be replaced by holes carefully made in plastic through which passes a metallic rod.}
\label{fig:figure2}
\end{figure}

Once the main components are printed, the second part of the construction process starts. The plastic elements printed are combined with components like nuts, screws, bolts, washers, springs and rods, easily found on a hardware store, to create a fully functional opto-mechanical component. Figures \ref{fig:figure2}A and \ref{fig:figure2}B depict a drive screw mechanism implemented by embedding a nut in the plastic. Figure \ref{fig:figure2}C shows the implementation of a linear bearing using a rod. It is interesting to note that even though the individual components added in the second part are not designed for high precision applications, we have found that the opto-mechanical components fabricated with them, provide very similar performance with respect to its commercial counterpart, as is shown in the results section.

\subsection*{Kinematic Mount / Kinematic Platform}
\begin{figure}[!h]
\centering
\subfloat[]
{\label{fig:figure1a}\includegraphics[width=.3\textwidth]{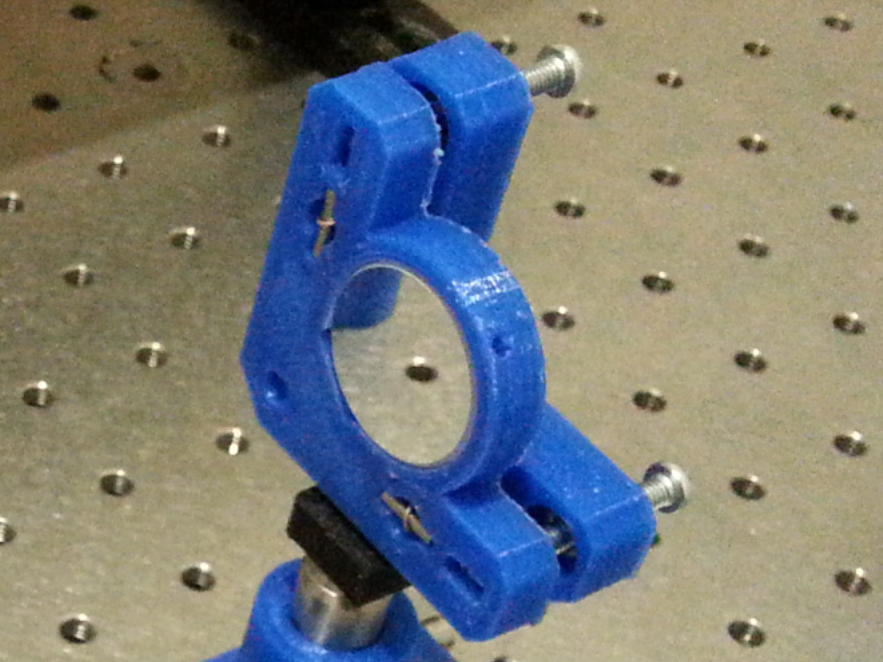}} \quad
\subfloat[]
{\label{fig:figure1b}\includegraphics[width=.3\textwidth]{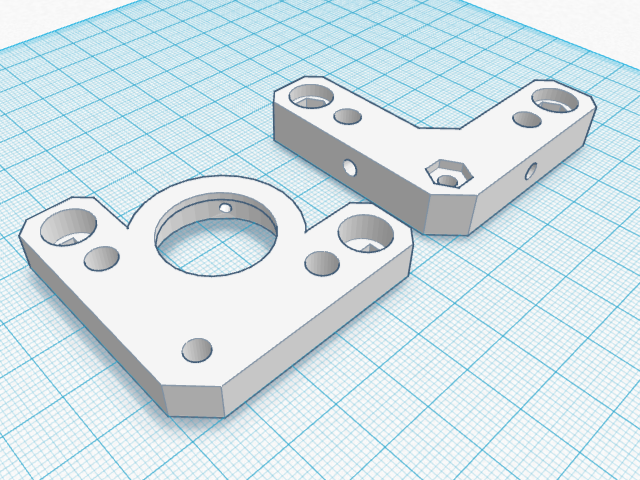}} \quad
\subfloat[]
{\label{fig:figure1c}\includegraphics[width=.3\textwidth]{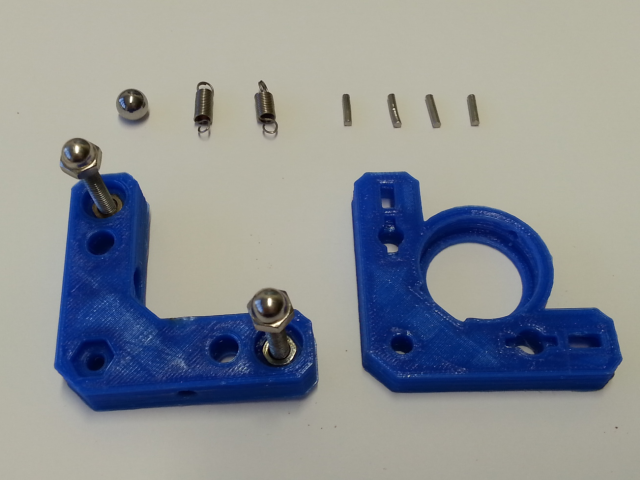}} \\
\caption{{\bf 3D printed kinematic mount.}
(A) Finished kinematic mount with mirror mounted. (B) 3D model of plastic components (KM top: leftmost component, KM bottom: rightmost component). (C) Components required to build the kinematic mount.}
\label{fig:figure3}
\end{figure}

A kinematic mount (KM) is an opto-mechanical component used to adjust precisely the tip and tilt of a mirror (or lens), while it holds the component securely in place, as shown in figure \ref{fig:figure3}A. Similarly, a kinematic platform (KP) may be seen as a rotated kinematic mirror mount that is mainly used to control the tip and tilt of a flat surface where other components such as prisms, beam splitters or non-standard optics are secured.

% Place figure captions after the first paragraph in which they are cited.

Our proposed implementation of the kinematic mirror mount, follows the widely used cone, groove, and flat constraint scheme\cite{Newport}, and is based on the original design of Doug Marett\cite{DougMarett} that can be found on Thingiverse. The mount is implemented by printing two pieces of plastic (figure \ref{fig:figure3}B) that are joint together using a sphere and two springs secured with two rods on each side. The drive screw mechanism is built using two nuts that are embedded into the plastic. Two M4 screws with rounded nuts in one end are used to adjust the tip and tilt respectively (figure \ref{fig:figure3}C). The rounded nuts are used to keep the two plastic pieces into position and to reduce any unwanted motion. A complete list of the required materials is presented in table~\ref{table1}.

% Place tables after the first paragraph in which they are cited.
\begin{table}[!ht]
\centering
\caption{
{\bf Bill of Materials for Kinematic Mount}}
\begin{tabular}{|l|l|l|l|}
\hline
{\bf Component} & {\bf Comments} & {\bf Quantity}& {\bf Unit cost [EUR]}\\ \hline\hline
KM top$^{\dag}$ & Vol 9.663 cm$^3$ & 1 & 4.8 \\ \hline
KM bottom$^{\dag}$ & Vol 10.588 cm$^3$ & 1 & 5.3 \\ \hline
Steel sphere & $\phi=8\,\mathrm{mm}$ & 1 & 0.25 \\ \hline
Hex nut & M4 & 4 & 0.25 \\ \hline
Rounded nut & M4 & 2 & 0.25 \\ \hline
Screw & M4, $L=4$ cm & 2 & 0.25 \\ \hline
Tension spring & $L=1$ cm & 2 & 0.25 \\ \hline
Metal rod & $L\approx1$ cm & 2 & 0.25  \\ \hline\hline
 & & {\bf TOTAL} & {\bf 13.35}  \\ \hline
\end{tabular}
\begin{flushleft} $^{\dag}$ Printing cost of 0.5 EUR / cm$^3$ is assumed. 
\end{flushleft}
\label{table1}
\end{table}

Typically, a mirror mount may cost between $35\euro{}$, for a basic mount, to $150\euro{}$, for a more advanced component. The values indicated correspond to an average cost over very popular manufacturers such as Newport, Thorlabs and Edmund Optics. Regarding the lead time, it may vary between 2-7 days for Europe or United States to 1 to 6 months for other countries (for instance 6 months for Colombia). 

It is clear that with plastic we cannot compete against high-performance components, due to the plastic mechanical properties; however, we can still reproduce the behaviour of low-end opto-mechanical components at a fraction of the cost and with a significant reduction in lead time to a few hours. 

The kinematic mount shown in figure \ref{fig:figure3}A was built in three hours: the first two hours were spent printing the two plastic components, and the last hour was spent building, adjusting and testing the component. The total cost of manufacture is $12\,\euro{}$, including the components purchased in a hardware store. As a result the cost is significantly reduced, but more importantly the lead time is dramatically reduced from months, to hours.

\subsection*{Translation stage}
\begin{figure}[!h]
\centering
\subfloat[]
{\label{fig:figure1a}\includegraphics[width=.3\textwidth]{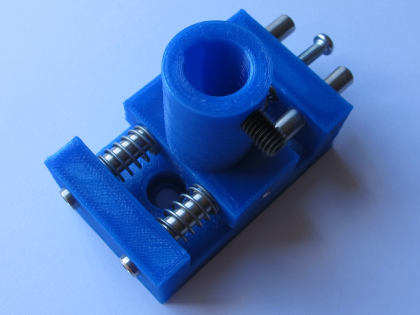}} \quad
\subfloat[]
{\label{fig:figure1b}\includegraphics[width=.3\textwidth]{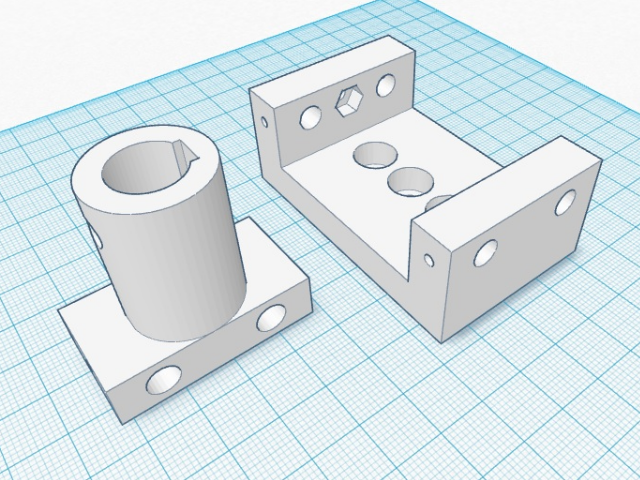}} \quad
\subfloat[]
{\label{fig:figure1c}\includegraphics[width=.3\textwidth]{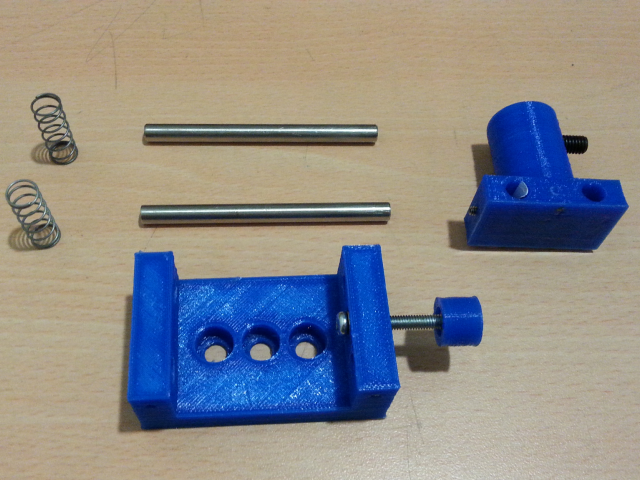}} \\
\caption{{\bf 3D printed translation stage.}
(A) Finished translation stage. (B) 3D model of plastic components (TS top: leftmost component, TS bottom: rightmost component). (C) Components required to build a translation stage.}
\label{fig:figure4}
\end{figure}

A translation stage (TS) is a component typically used to vary the position of an object along a single axis. The position of the moving platform is controlled with a drive screw mechanism more precise that the one used in the kinematic mount (figure \ref{fig:figure4}A).

The translation stage is composed of the two pieces shown in figure \ref{fig:figure4}B. Unlike a traditional system (for instance see ref~\cite{Giddyup9}), where the rods are fixed and the moving section has two linear bearings; the stage printed has the two rods fixed to the moving section, and the walls of the main platform act as bearings.
With this scheme it was possible to obtain a very stable and robust linear motion along the axis of the rods. The drive screw mechanism is built using a single nut that is embedded in one wall of the main platform. A M4 screw with a rounded nut in one end is used to set the position of the moving section over a range of $1.0\,\mathrm{cm}$.

According to our experience, a standard translation stage cannot be found for less than $150\euro{}$. This amount corresponds to a considerable investment for a laboratory or company that is on an early stage of development. Regarding the lead time, there is no much difference with respect to the kinematic mount presented above.

The component shown in figure \ref{fig:figure4}A was built in four hours. Three hours to print and clean the plastic components and one hour to leave the stage fully operational. In table \ref{table2} are indicated the required materials and the corresponding manufacturing costs.

% Place tables after the first paragraph in which they are cited.
\begin{table}[!ht]
\centering
\caption{
{\bf Bill of Materials for Kinematic Mount}}
\begin{tabular}{|l|l|l|l|}
\hline
{\bf Component} & {\bf Comments} & {\bf Quantity}& {\bf Unit cost [EUR]}\\ \hline\hline
TS top$^{\dag}$ & Vol 14.848 cm$^3$ & 1 & 7.4 \\ \hline
TS bottom$^{\dag}$ & Vol 9.663 cm$^3$ & 1 & 16.3 \\ \hline
Steel rod & $\phi=4\,\mathrm{mm}$, $L\approx7.5\,\mathrm{cm}$ & 2 & 4 \\ \hline
Hex nut & M4 & 1 & 0.25 \\ \hline
Rounded nut & M4 & 1 & 0.25 \\ \hline
Screw & M4, $L=4$ cm & 1 & 0.25 \\ \hline
Grub screw & M6, $L=1$ cm & 1 & 0.25 \\ \hline
Spring & $L=1.5$ cm & 2 & 0.25 \\ \hline
 & & {\bf TOTAL} & {\bf 33.2}  \\ \hline
\end{tabular}
\begin{flushleft} $^{\dag}$ Printing cost of 0.5 EUR / cm$^3$ is assumed.
\end{flushleft}
\label{table2}
\end{table}

%\subsection*{Post Holders and Post Clamps}

\subsection*{Integrating sphere}
An integrating sphere is an optical component composed of a hollow spherical cavity that has two small windows orientated at $90^{\circ}$ one with respect to the other~\cite{Jacqueza}. The first window corresponds to the input port whereas the other is the output port where a detector is located. For commercial devices, the interior of the sphere is covered with a diffuse reflective coating. When light enters to the sphere, it is reflected equally in all directions due to scattering. As a result, an integrating sphere is a device that can be used to measure optical power, while the spatial information of the input beam is destroyed (i.e. beam shape and entrance angle).

\begin{figure}[!h]
\centering
\subfloat[]
{\label{fig:figure1a}\includegraphics[width=.3\textwidth]{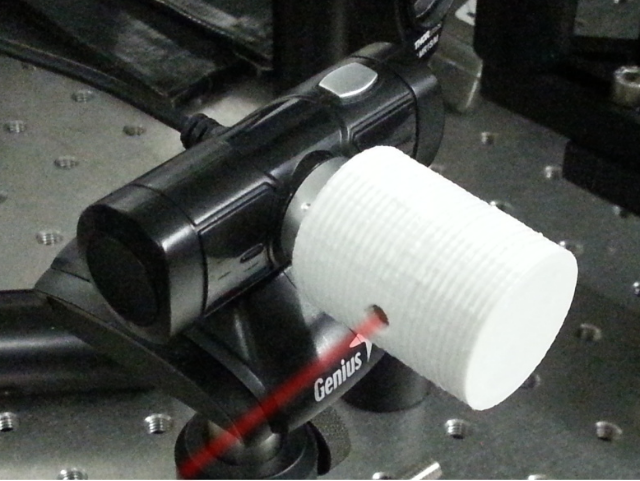}} \quad
\subfloat[]
{\label{fig:figure1b}\includegraphics[width=.3\textwidth]{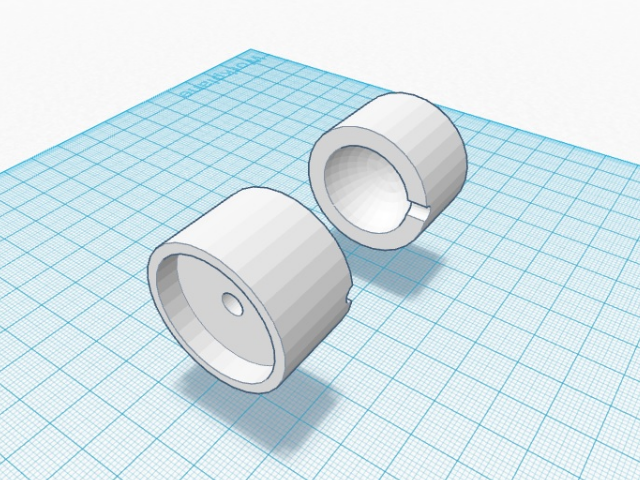}} \quad
\subfloat[]
{\label{fig:figure1c}\includegraphics[width=.3\textwidth]{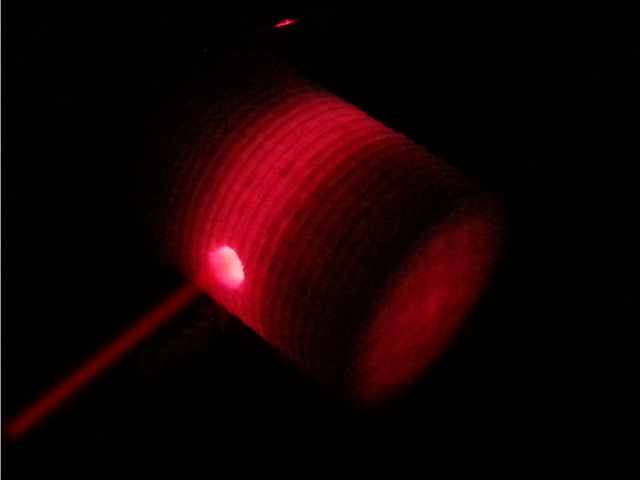}} \\
\caption{{\bf 3D printed integrating sphere.}
Panel (a). Integrating sphere using a webcam as a detector.  Panel (b). 3D model of the plastic integrating sphere. The inner sphere is printed as a complete piece. In the figure the sphere is divided in two sections for ilustrative purposes. Panel (c). Integrating sphere operating with lights off.}
\label{fig:figure5}
\end{figure}

In Fig.~\ref{fig:figure5} is shown the 3D printed integrating sphere developed. The hollow sphere is printed using white PLA as shown in figure \ref{fig:figure5}B. In our design the output window allows to use any of the following detectors: photodiode, webcam, or the camera of a mobile phone. In the case of using a webcam, a program written in Python is used to calibrate the integrating sphere with respect to a known reference and also to perform further power measurements.

For each image taken with the webcam, a corresponding light input power is determined by calculating the area behind the grayscale histogram of a masked version of the image. The mask used is a circular mask of radius 100 pixels. In our setup we used a Genius Eye 100 webcam.

In table~\ref{table3} is presented the cost of fabricating the integrating sphere shown in figure \ref{fig:figure5}. The coupler characteristics depends on the detector used.

% Place tables after the first paragraph in which they are cited.
\begin{table}[!ht]
\centering
\caption{
{\bf Bill of Materials for Kinematic Mount}}
\begin{tabular}{|l|l|l|l|}
\hline
{\bf Component} & {\bf Comments} & {\bf Quantity}& {\bf Unit cost [EUR]}\\ \hline\hline
IS top$^{\dag}$ & Vol 20.475 cm$^3$ & 1 & 10.2 \\ \hline
IS coupler$^{\dag}$ & Vol 1.6 cm$^3$ & 1 & 0.8 \\ \hline

 & & {\bf TOTAL} & {\bf 11}  \\ \hline
\end{tabular}
\begin{flushleft} $^{\dag}$ Printing cost of 0.5 EUR / cm$^3$ is assumed.
\end{flushleft}
\label{table3}
\end{table}

% Results and Discussion can be combined.
\section*{Results and Discussion}
\begin{figure}[!h]
\centering
\subfloat[]
{\label{fig:figure1a}\includegraphics[width=.45\textwidth]{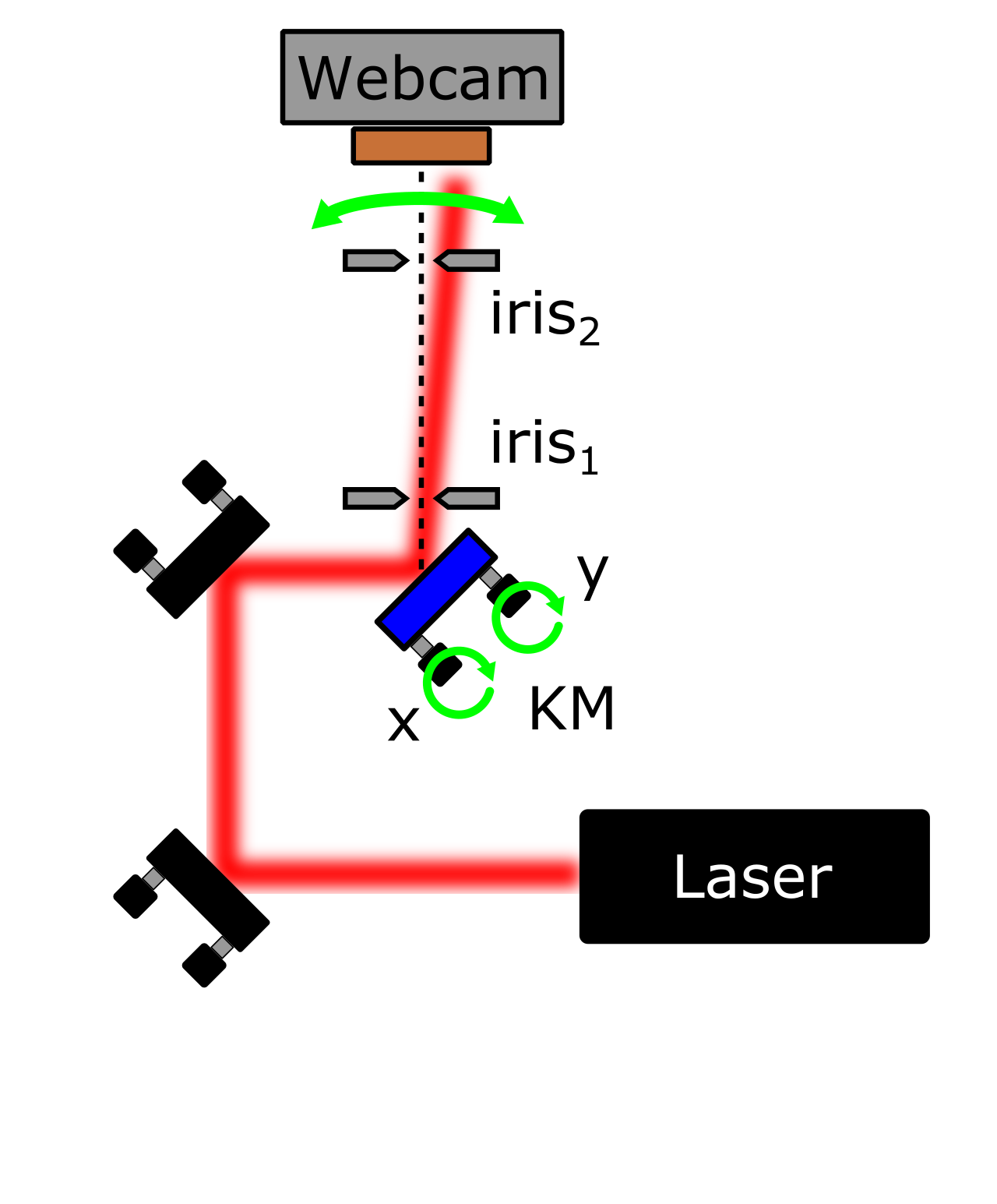}} \quad
\subfloat[]
{\label{fig:figure1c}\includegraphics[width=.45\textwidth]{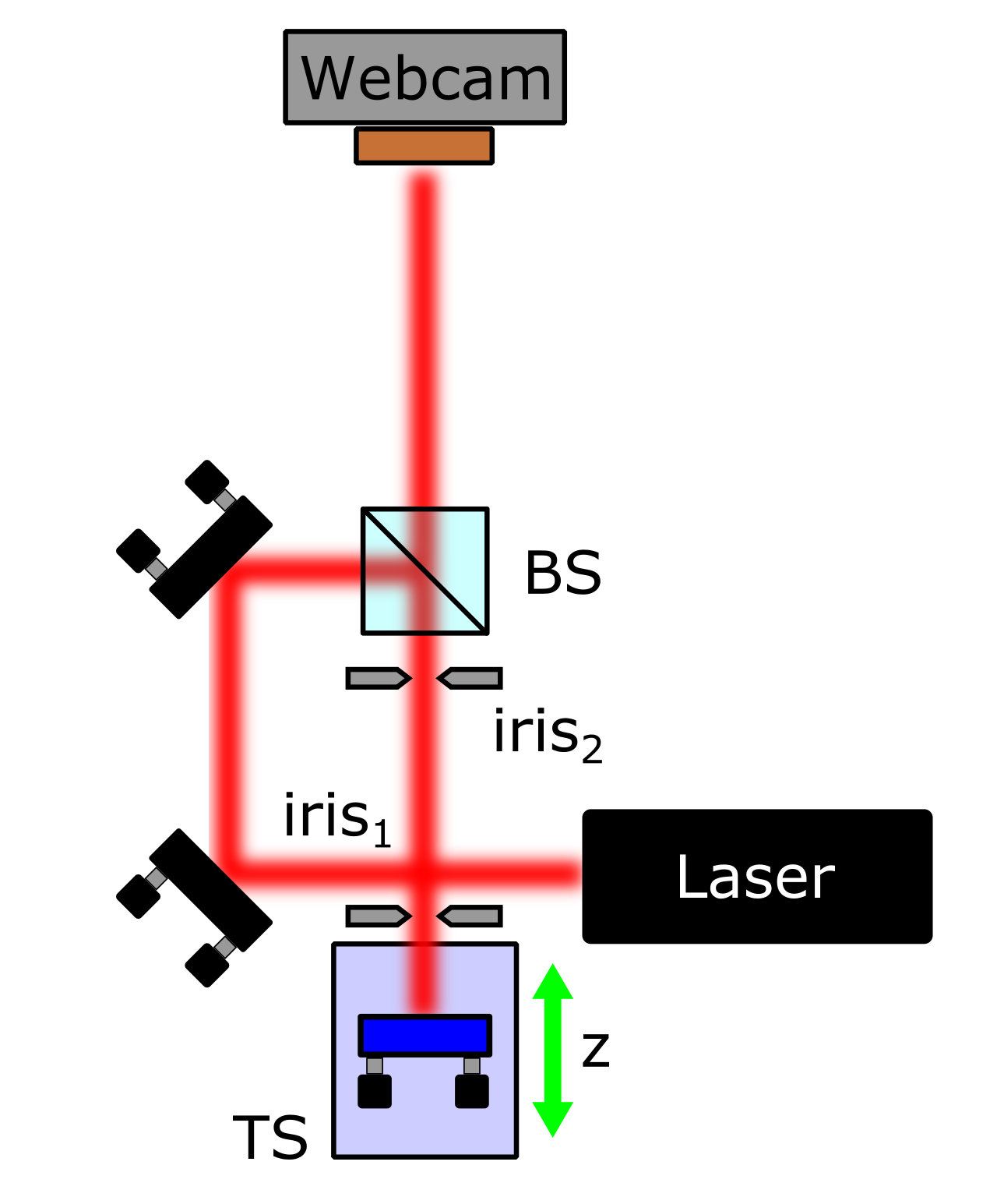}} \\
\caption{{\bf Experimental setup used to characterize 3D printed components.} (A) Setup used to compare two kinematic mounts; KM kinematic mount under test. (B) Setup used to compare two translation stages; TS translation stage under test.}
\label{fig:figure6}
\end{figure}

To determine the performance of the opto-mechanical components fabricated, a direct comparison with respect to its commercial counterpart was carried out by using the experimental setups presented in Fig. \ref{fig:figure6}.

In Fig. \ref{fig:figure6}A it is shown the scheme used to compare the kinematic mounts. The input beam is generated using a He-Ne laser with Gaussian spatial profile and a beam waist of $\sim 1\,\mathrm{mm}$. After two reflections, the beam is reflected in a mirror mounted on the kinematic mount to be tested (indicated in blue) and its centroid position is monitored using a webcam located at $80\,\mathrm{cm}$ with respect to the mirror vertical axis. A routine written in Python, that uses the OpenCV library, records the beam centroid as a function of different angles. The beam reference position is determined by aligning initially the beam with respect to the two irises 1 and 2 located before the camera.

To test the performance of the translation stage, the setup shown in Fig. \ref{fig:figure6}B is used. The beam reflected by the beam splitter (BS) is reflected by a mirror located in the translation stage to be characterized. The TS is positioned in such a way that the beam reflected passes through the two irises for different positions.

In order to evaluate the performance of each opto-mechanical component, two sets of measurements were carried out. The first is taken using the 3D printed component and the second the commercial device. 

For the kinematic mount, the interval $(-1\,\mathrm{^\circ}, +1\,\mathrm{^\circ})$ in the horizontal and vertical directions is divided equally in seven points. Each direction is scanned ten times by rotating either the X and Y knob, in order to evaluate the hysteresis and repetability of the component. 

\begin{figure}[!h]
\centering
\subfloat[]
{\label{fig:figure1a}\includegraphics[width=.45\textwidth]{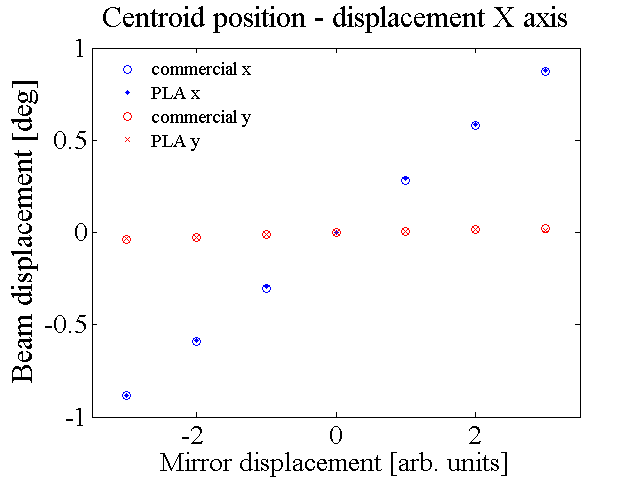}} \quad
\subfloat[]
{\label{fig:figure1c}\includegraphics[width=.45\textwidth]{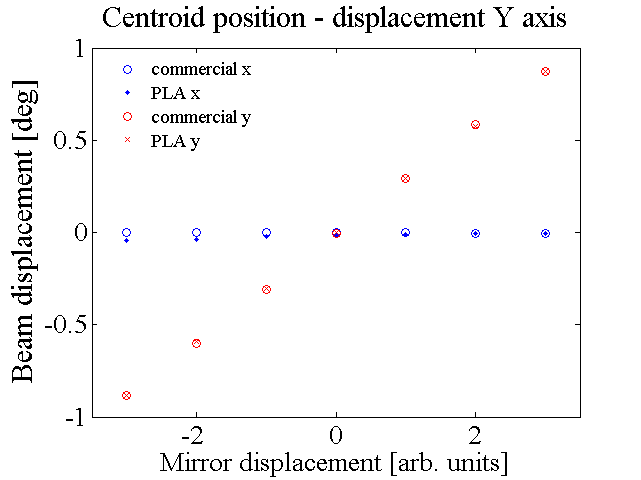}} \\
\caption{{\bf Experimental results kinematic mount.} Centroid position as a function of the X knob (A) and Y knob (B).}
\label{fig:figure7}
\end{figure}

Figures \ref{fig:figure7}A and \ref{fig:figure7}B show the centroid position as a function of the X or Y knob rotation, respectively. In both cases is clearly seen that the 3D printed kinematic mount, exhibits the same behaviour as its commercial counterpart, in this case a Thorlabs KM100 mount. In fact, when the beam is shifted in one direction, the other is confined within the same small interval as the commercial component. From the characterization, the only difference observed between both components appears in the sensitivity experienced in the knob rotation, determined by the screw thread.

\begin{figure}[!h]
\centering
\includegraphics[width=.45\textwidth]{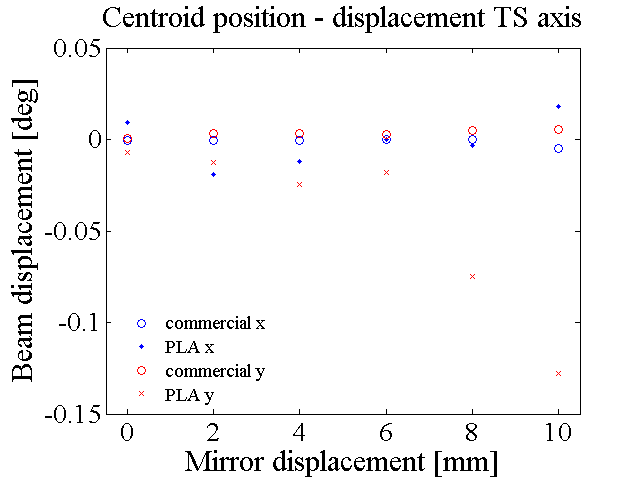}
\caption{{\bf Experimental results translation stage.} Centroid position as a function of the mirror displacement in millimeters.}
\label{fig:figure8}
\end{figure}

For the translation stage, two set of measurements where performed over the interval $(0\,\mathrm{mm}, +10\,\mathrm{mm})$, where the translation step was set to $2\,\mathrm{mm}$. The results are shown in Fig. \ref{fig:figure8}. From the results is observed for the commercial TS, the beam drift lies within the interval $(-0.01\,\mathrm{^\circ}, +0.01\,\mathrm{^\circ})$. On the other hand, the 3D printed TS provides a worse performance, particularly for displacements beyond $6\,\mathrm{mm}$, where the rounded screw exerts a significant pressure on the moving platform. This pressure gives rise to the unwanted displacement observed in the y-direction. 

Fortunately, for small translations, below $6\,\mathrm{mm}$, the device presents an acceptable performance where the beam experiences a drift that lies within an interval of tenths of degrees. Notice that this unwanted beam displacement is imperceptible to the eye.

\begin{figure}[!h]
\centering
\subfloat[]
{\label{fig:figure1a}\includegraphics[width=.45\textwidth]{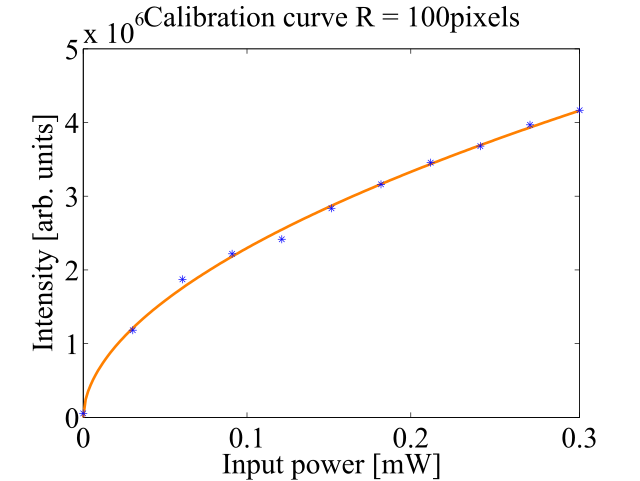}} \quad
\subfloat[]
{\label{fig:figure1c}\includegraphics[width=.45\textwidth]{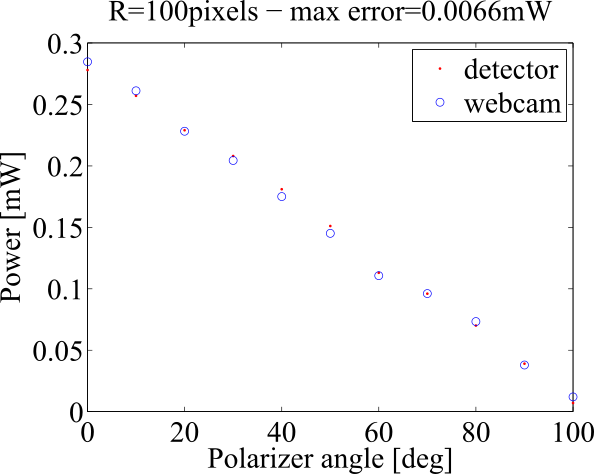}} \\
\caption{{\bf Experimental results integrating sphere.} (A) device response as a function of the intensity (area behind image histogram). (B) Comparison with a commercial power meter after calibration.}
\label{fig:figure9}
\end{figure}

Regarding the integrating sphere, it is found that it exhibits a non linear response as a function of the input beam intensity. Figure~\ref{fig:figure9}A displays an example of the device response as a function of the input intensity. From the fitted data, a calibration curve is obtained that is further used to determine the response of the power meter. For the sake of example, Fig.~\ref{fig:figure9}B presents a single wavelength ($\lambda=633\,\mathrm{nm}$) comparison between a commercial power meter and a 3D printed integrating sphere. From the data is observed a maximum relative error of less than $2\%$. In the experiment the light intensity is controlled by changing the angle between two crossed polarizers.

\section*{Conclusion}
We have developed and characterized a set of opto-mechanical components that can be easily implemented using a 3D printer based on Fused Filament Fabrication (FFF) and parts that can be found on any hardware store. In particular we have compared three of the main components required to implement a Michelson-Morley interferometer, namely a kinematic mount, a translation stage and an integrating sphere with respect to commercial alternatives.

From our results, we have found that 3D printing provides a suitable alternative to implement experimental equipment in scenarios where is not required a high precision. Surprisingly, the 3D printed kinematic mount provides a very similar performance with respect to its commercial counterpart. Even though the results obtained for the translation stage are not so optimistic, since the beam drift is imperceptible to the eye, the device can be suitable for undergraduate laboratories. Regarding the integrating sphere, we have developed and demonstrated a simple accessory that can be printed in order to convert a webcam into a power detector.

Importantly, in all cases we have found that a 3D printer is an extremely useful resource in any laboratory since it opens the possibility to fabricate experimental equipment that is highly customizable, at a low cost with respect to commercial alternatives, and more importantly in a very small period of time.

%\section*{Supporting Information}

% Include only the SI item label in the paragraph heading. Use the \nameref{label} command to cite SI items in the text.
%\paragraph*{S1 File.}
%\label{S1_File}
%{\bf Lorem Ipsum.}  Maecenas convallis mauris sit amet sem ultrices gravida. Etiam eget sapien nibh. Sed ac ipsum eget enim egestas ullamcorper nec euismod ligula. Curabitur fringilla pulvinar lectus consectetur pellentesque.

\section*{Acknowledgments}
LJSS and AV acknowledge support from Facultad de Ciencias, U. de Los Andes.

%*** \nolinenumbers ***

% Either type in your references using
% \begin{thebibliography}{}
% Text
% \end{thebibliography}
%
% or
%
% Compile your BiBTeX database using our plos2015.bst
% style file and paste the contents of your .bbl file
% here.
% 


\begin{thebibliography}{10}
\bibitem{Economist} P. Markillie P, ``A third industrial revolution'', The Economist, 21 Apr 2012

\bibitem{Berman} B. Berman, ``3D printing: The new industrial revolution'', Business Horizons, 55, 155 (2012)

\bibitem{Pearce2013a} B. T. Wittbrodt, A. G. Glover, J. Laureto, G. C. Anzalone, D. Oppliger, J. L. Irwin, J. M. Pearce, ``3D printing: The new industrial revolution'', Mechatronics, 23, 713 (2013)

\bibitem{Pearce_book} J. M. Pearce, ``Open-Source Lab: How to Build Your Own Hardware and Reduce Research Costs'', Amsterdam, Elsevier, (2014) 

\bibitem{Baden} T. Baden, A.M. Chagas, G. Gage, T. Marzullo, LL. Prieto-Godino, T. Euler, ``Open Labware: 3-D Printing Your Own Lab Equipment'', PLoS Biol 13(3): e1002086 (2015)

\bibitem{PearceScience} J. M. Pearce, ``Building Research Equipment with Free, Open-Source Hardware'',  Science 337: 1303–1304 (2012)

\bibitem{Pearce2013b} C. Zhang, N. C. Anzalone, R. P. Faria, J. M. Pearce, ``Open-Source 3D-Printable Optics Equipment'', PLoS ONE 8(3): e59840 (2013) 

\bibitem{Pearce2014} B. Wijnen, E. J. Hunt, G. C. Anzalone, J. M. Pearce, ``Open-Source Syringe Pump Library'', PLoS ONE 9(9): e107216 (2014)

\bibitem{Zavattieri} W. Gao, Y. Zhang, D. Ramanujan, K. Ramani, Y. Chen, C. B. Williams, C. C. L. Wang, Y. C. Shin, S. Zhang, P. D. Zavattieri, ``The status, challenges, and future of additive manufacturing in engineering'', Computer-Aided Design, 69, 65 (2015)

\bibitem{Piller} C. Weller, R. Kleer and F. T. Piller, ``Economic implicationsof 3D printing: Market structure models in light of additive manufacturing revisited'', Int. J. Production Economics, 164, 43 (2015)

\bibitem{Euler} T. Baden, A. M. Chagas, G. Gage, T. Marzullo, L. L. Prieto-Godino and T. Euler, ``Open Labware: 3-D Printing Your Own Lab Equipment'', PLoS Biol 13, 3 (2015) 

\bibitem{PrusaTayrona} Prusa Tayrona 3D printer. Availale: http://make-r.co [Accessed 2016 Jun 22]

\bibitem{Newport} Technical note: Optical Mirror Mount Technology Guide. Available: https://www.newport.com/optical-mirror-mount-technology-guide [Accessed 2016 Jun 22]

\bibitem{DougMarett} Optical mount for lens / mirror / prism. Available: http://www.thingiverse.com/thing:1004337 [Accessed 2016 Jun 22]

\bibitem{Giddyup9} Short travel manual translation stage. Available: http://www.thingiverse.com/thing:151481 [Accessed 2016 Jun 22]

\bibitem{Jacqueza} J. A. Jacqueza, H. F. Kuppenheim, ``Theory of the integrating sphere'', J.Opt.Soc.Am., 45, 460 (1955)

\end{thebibliography}
\end{document}